\documentclass[longauth]{aa}
\usepackage{graphicx}
\usepackage{txfonts}
\usepackage{xcolor}
\usepackage{hyperref}
\hypersetup{colorlinks, linkcolor=blue, citecolor=blue, urlcolor=blue}
\usepackage{amsmath}
\usepackage{ulem}
\usepackage{soul}
\usepackage{enumitem}

\newcommand{\pivec}{\mbox{\boldmath $\pi$}}

\newcommand{\te}{t_{\rm E}}
\newcommand{\thetae}{\theta_{\rm E}}

\newcommand{\pie}{\pi_{\rm E}}
\newcommand{\pien}{\pi_{{\rm E},N}}
\newcommand{\piee}{\pi_{{\rm E},E}}
\newcommand{\dl}{D_{\rm L}}
\newcommand{\ds}{D_{\rm S}}

\definecolor{brown}{rgb}{0.59, 0.29, 0.0}
\definecolor{darkgreen}{rgb}{0.0, 0.42, 0.24}
\definecolor{darkblue}{rgb}{0.01, 0.31, 0.59}
\definecolor{darkblue}{rgb}{0.0, 0.25, 0.42}
\definecolor{blue}{rgb}{0.0,0.0,1.0}
\definecolor{green}{rgb}{0.0,1.0,0.0}

\begin{document}

\title{KMT-2021-BLG-2010Lb, KMT-2022-BLG-0371Lb, and KMT-2022-BLG-1013Lb: Three microlensing planets
detected via partially covered signals}
\titlerunning{Microlensing planets detected from partially covered signals}

\author{
     Cheongho~Han\inst{01} 
\and Chung-Uk~Lee\inst{02} 
\and Weicheng~Zang\inst{03,04}     
\and Youn~Kil~Jung\inst{02} 
\and Grant W. Christie\inst{05}
\and Jiyuan Zhang\inst{04}
\\
(Leading authors)\\
     Michael~D.~Albrow\inst{06}   
\and Sun-Ju~Chung\inst{02}      
\and Andrew~Gould\inst{07,08}      
\and Kyu-Ha~Hwang\inst{02} 
\and Doeon~Kim\inst{01}
\and Yoon-Hyun~Ryu\inst{02} 
\and In-Gu~Shin\inst{03} 
\and Yossi~Shvartzvald\inst{09}   
\and Hongjing~Yang\inst{04}     
\and Jennifer~C.~Yee\inst{03}   
\and Sang-Mok~Cha\inst{02,10} 
\and Dong-Jin~Kim\inst{02} 
\and Seung-Lee~Kim\inst{02,11} 
\and Dong-Joo~Lee\inst{02} 
\and Yongseok~Lee\inst{02,10} 
\and Byeong-Gon~Park\inst{02} 
\and Richard~W.~Pogge\inst{05}
\\
(The KMTNet Collaboration)\\
     Tim Natusch\inst{05,12}
\and Shude Mao\inst{04,13}
\and Dan Maoz\inst{14}
\and Matthew T. Penny\inst{15} 
\and Wei Zhu\inst{04}
\\
(The MAP \& $\mu$FUN Follow-up Teams)
}

\institute{
     Department of Physics, Chungbuk National University, Cheongju 28644, Republic of Korea  \\ \email{cheongho@astroph.chungbuk.ac.kr}    
\and Korea Astronomy and Space Science Institute, Daejon 34055, Republic of Korea                                                          
\and Center for Astrophysics $|$ Harvard \& Smithsonian 60 Garden St., Cambridge, MA 02138, USA                                            
\and Department of Astronomy and Tsinghua Centre for Astrophysics, Tsinghua University, Beijing 100084, China                              
\and Auckland Observatory, Auckland, New Zealand                                                                                           
\and University of Canterbury, Department of Physics and Astronomy, Private Bag 4800, Christchurch 8020, New Zealand                       
\and Max Planck Institute for Astronomy, K\"onigstuhl 17, D-69117 Heidelberg, Germany                                                      
\and Department of Astronomy, The Ohio State University, 140 W. 18th Ave., Columbus, OH 43210, USA                                         
\and Department of Particle Physics and Astrophysics, Weizmann Institute of Science, Rehovot 76100, Israel                                 
\and School of Space Research, Kyung Hee University, Yongin, Kyeonggi 17104, Republic of Korea                                             
\and Korea University of Science and Technology, 217 Gajeong-ro, Yuseong-gu, Daejeon, 34113, Republic of Korea                             
\and Institute for Radio Astronomy and Space Research (IRASR), AUT University, Auckland, New Zealand                                       
\and National Astronomical Observatories, Chinese Academy of Sciences, Beijing 100101, China                                               
\and School of Physics and Astronomy, Tel-Aviv University, Tel-Aviv 6997801, Israel                                                        
\and Department of Physics and Astronomy, Louisiana State University, Baton Rouge, LA 70803 USA                                            
}
\date{Received ; accepted}

\abstract
{}
{
We inspect 4 microlensing events KMT-2021-BLG-1968, KMT-2021-BLG-2010, KMT-2022-BLG-0371, and 
KMT-2022-BLG-1013, for which the light curves exhibit partially covered short-term central 
anomalies. We conduct detailed analyses of the events with the aim of 
revealing the nature of the anomalies. 
}
{
We test various models that can give rise to the anomalies of the individual events including 
the binary-lens (2L1S) and binary-source (1L2S) interpretations. Under the 2L1S interpretation, 
we thoroughly inspect the parameter space to check the existence of degenerate solutions, and 
if they exist, we test the feasibility of resolving the degeneracy.
}
{
We find that the anomalies in KMT-2021-BLG-2010 and KMT-2022-BLG-1013 are uniquely defined by 
planetary-lens interpretations with the planet-to-host mass ratios of $q\sim 2.8\times 10^{-3}$ 
and $\sim 1.6\times 10^{-3}$, respectively.  For KMT-2022-BLG-0371, a planetary solution with 
a mass ratio $q\sim 4\times 10^{-4}$ is strongly favored over the other three degenerate 2L1S 
solutions with different mass ratios based on the $\chi^2$ and relative proper motion arguments, 
and a 1L2S solution is clearly ruled out.  For KMT-2021-BLG-1968, on the other hand, we find 
that the anomaly can be explained either by a planetary or a binary-source interpretation, 
making it difficult to firmly identify the nature of the anomaly.  From the Bayesian analyses 
of the identified planetary events, we estimate that the masses of the planet and host are 
$(M_{\rm p}/M_{\rm J}, M_{\rm h}/M_\odot) = 
 (1.07^{+1.15}_{-0.68}, 0.37^{+0.40}_{-0.23})$, 
$(0.26^{+0.13}_{-0.11}, 0.63^{+0.32}_{-0.28})$, and
$(0.31^{+0.46}_{-0.16}, 0.18^{+0.28}_{-0.10})$ 
for KMT-2021-BLG-2010L, KMT-2022-BLG-0371L, and KMT-2022-BLG-1013L, respectively.
}
{}


\maketitle

\section{Introduction}\label{sec:one}

Current planetary microlensing experiments are being carried out with the use of multiple wide field
telescopes. With the high observational cadence achieved by employing large-format cameras and
continuous coverage using multiple telescopes, the experiments aim to construct a large planet
sample including terrestrial planets. The microlensing planet sample is of scientific importance 
in studying the demographic distribution of planets because it includes planet populations to which
the detection efficiency of other major planet detection methods is low, for example, cold planets
with faint host stars.

Despite the enhanced observational cadence of the current microlensing surveys, the coverage for 
a fraction of planet-induced signals can be incomplete, especially for those with short durations. 
The most common cause for the incomplete coverage is a bad weather. In the case of the Korea 
Microlensing Telescope Network \citep[KMTNet:][]{Kim2016} experiment, its three telescopes are 
located in the three continents of the Southern Hemisphere including the Siding Spring Observatory 
in Australia (KMTA), the Cerro Tololo Interamerican Observatory in Chile (KMTC), and the South 
African Astronomical Observatory in South Africa (KMTS). If one of these telescope sites is 
clouded out, the coverage of a signal with a duration $\lesssim  1$~day produced by a low-mass 
planet would be incomplete. Another cause for the incomplete anomaly coverage is that a subset 
of survey fields are observed with a relatively low cadence. The observational cadence for the 
peripheral fields of the KMTNet survey is 4--20 times lower than the 0.25~hr cadence of the prime 
fields, and thus the coverage of a short-duration signal detected in these low-cadence fields 
can be incomplete, for example, the planetary events KMT-2017-BLG-0673 and KMT-2019-BLG-0414 
\citep{Han2022a}. Besides, the coverage of signals detected in the early or late phase of a bulge 
season can also be incomplete, for example, the two-planet event OGLE-2019-BLG-0468 \citep{Han2022b}, 
due to fact that the target, respectively, rises and sets close to twilight.

In this paper, we present the analyses of 4 microlensing events KMT-2021-BLG-1968,
KMT-2021-BLG-2010, KMT-2022-BLG-0371, and KMT-2022-BLG-1013, for which the light curves
of the events exhibit partially covered short-term central anomalies. With the aim of
revealing the nature of the anomalies, we conduct detailed analyses of the events under 
various interpretations that can give rise to the anomalies of the individual events.

We present the analyses of the events according to the following organization. In 
Sect.~\ref{sec:two}, we describe the acquisition and reduction procedure of the data used in 
the analyses. In Sect.~\ref{sec:three}, we depict the models tested for the interpretations 
of the anomalies, and introduce parameters used in the modeling. In the subsequent subsections, 
we present the detailed analyses conducted for the events KMT-2021-BLG-1968 
(Sect.~\ref{sec:three-one}), KMT-2021-BLG-2010 (Sect.~\ref{sec:three-two}), KMT-2022-BLG-0371 
(Sect.~\ref{sec:three-three}), and KMT-2022-BLG-1013 (Sect.~\ref{sec:three-four}). Based on 
these analyses, we judge whether the anomalies are of planetary origin or not. In 
Sect.~\ref{sec:four}, we specify the source stars of the events by estimating the colors and 
magnitudes, and estimate angular Einstein radii. In Sect.~\ref{sec:five}, we estimate the 
physical parameters of the mass and distance to the identified planetary system. We summarize 
results in Sect.~\ref{sec:six}.

\section{Observations and data}\label{sec:two}

All the lensing events analyzed this work were detected in the 2021 and 2022 seasons by the
KMTNet survey from the monitoring of stars lying toward the Galactic bulge field. The three 
telescopes used by the KMTNet survey are identical, and each 1.6~m telescope is equipped with 
a camera yielding 4~deg$^2$ field of view.  Since July 2020, the MAP \& $\mu$FUN Follow-up 
teams have been conducting a long-term follow-up program for high-magnification events 
\citep{Zang2021}.  For the two very high-magnification events KMT-2022-BLG-0371 and 
KMT-2022-BLG-1013, observations were additionally conducted using the 1~m telescopes of the 
Las Cumbres Observatory global network at CTIO (LCOC) and SAAO (LCOS) for both KMT-2022-BLG-0371 
and KMT-2022-BLG-1013, and using the 0.4~m telescope at the Auckland Observatory in New Zealand 
for KMT-2022-BLG-1013.  KMTNet also went into "followup mode" and increased its observational 
cadence over the peaks of these two high-magnification events.  Images from the KMTNet survey 
and MAP followup observations were mostly acquired in the $I$ band, and a fraction of KMTNet 
images were obtained in the $V$ band for the source color measurement.  The Auckland Observatory 
data were taken in the Wratten \#12 filter.

Image reductions and source star photometry of the KMTNet, LCO, and Auckland data were done
with the use of the pySIS pipeline developed by \citet{Albrow2009} and Yang et al. (in prep) 
on the basis of the difference image method \citep{Tomaney1996,Alard1998}. The error bars of 
data estimated by the individual photometry codes were readjusted following the \citet{Yee2012} 
routine in order that they are consistent with the scatter of data and $\chi^2$ per degree of 
freedom (dof) for each data set becomes unity.

\section{Analyses }\label{sec:three}

The anomalies in the lensing light curves of the events analyzed in this work were identified 
from the visual inspection of the microlensing data collected during the 2021 and 2022 seasons. 
The common characteristic of the events and the anomalies appearing in their lensing light curves 
is that the peak magnifications of the events are very high, with $A_{\rm peak}\sim 100$, 60, 
570, and 280 for KMT-2021-BLG-1968, KMT-2021-BLG-2010, KMT-2022-BLG-0371, and KMT-2022-BLG-1013, 
respectively, and the anomalies appear near the peaks of the light curves. There are three 
channels that can produce such central anomalies, including the planetary, binary-lens, and 
binary-source channels.

A central anomaly through the planetary channel arises when a source star passes through the
central anomaly region formed around the tiny caustic induced by a planetary companion to the
primary lens. A planetary companion induces two sets of caustics, in which one (central caustic)
forms near the position of the planet host \citep{Chung2005}, and the other (planetary caustic)
forms away from the host at a position $\sim {\bf s}-1/{\bf s}$ \citep{Han2006}, where ${\bf s}$ 
denotes the host-planet separation vector with its length scaled to the angular Einstein radius 
$\thetae$. Then, the chance for the peak region of a high-magnification event, resulting from 
the close approach of the source to the planet host, being perturbed by a planet is very high 
\citep{Griest1998}.

\begin{table*}[t]
\small
\caption{Model parameters of KMT-2021-BLG-1968 \label{table:one}}
\begin{tabular}{llll}
\hline\hline
\multicolumn{1}{c}{Parameter}        &
\multicolumn{1}{c}{2L1S (close)}       &
\multicolumn{1}{c}{2L1S (wide)}        &
\multicolumn{1}{c}{1L2S}        \\
\hline
$\chi^2$/dof              &  $7356.9/7360        $  &   $7357.8/7360        $   &  $7359.9/7360        $   \\
$t_0$ (HJD$^\prime$)      &  $9429.849 \pm 0.004 $  &   $9429.848 \pm 0.003 $   &  $9429.781 \pm 0.005 $   \\
$u_0$ ($10^{-3}$)         &  $10.15 \pm 0.54     $  &   $10.41 \pm 0.50     $   &  $11.10 \pm 0.52     $   \\
$t_{0,2}$ (HJD$^\prime$)  &  --                     &   --                      &  $9430.245 \pm 0.009 $   \\
$u_{0,2}$ ($10^{-3}$)     &  --                     &   --                      &  $4.36 \pm 0.41      $   \\
$\te$ (days)              &  $20.65 \pm 1.02     $  &   $20.27 \pm 0.93     $   &  $20.18 \pm 0.79     $   \\
$s$                       &  $0.618 \pm 0.017    $  &   $1.662 \pm 0.056    $   &  --                      \\
$q$ ($10^{-3}$)           &  $3.08 \pm 0.37      $  &   $3.14 \pm 0.47      $   &  --                      \\
$\alpha$ (rad)            &  $3.587 \pm 0.008    $  &   $3.592 \pm 0.012    $   &  --                      \\
$\rho$ ($10^{-3}$)        &  $< 4                $  &   $< 4                $   &  --                      \\
$\rho_2$ ($10^{-3}$)      &  --                     &   --                      &  --                      \\
$q_F$                     &  --                     &   --                      &  $0.123 \pm 0.015    $   \\
\hline                                                                                                                                                   
\end{tabular}
\end{table*}

A binary lens with roughly equal mass components can also induce a central anomaly. In the
binary case, a single caustic forms near the barycenter of a close binary lens with $s\ll 1$, 
while two caustics form near the positions of the individual lens components of a wide binary 
lens with $s\gg 1$. Then, the chance of the central perturbation of a high-magnification event 
resulting from the close approach of the source either to the barycenter of the close binary or 
to each of the wide lens components of a wide binary is high \citep{Han2009}.

A central anomaly can also arise when the closely spaced components of a binary source
successively approach the lens. In this case, the resulting light curve is the superposition of 
those of the events involved with the individual source stars, and the combined magnifications 
can appear to be anomalous.

We analyze the individual lensing events by modeling the light curves under the planetary, 
binary-lens, and binary-source interpretations. In each modeling, we search for a lensing solution, 
which specifies a set of lensing parameters depicting the lensing light curve. In the simplest case 
of a lensing event involving a single lens and a single source (1L1S), the lensing light curve is 
described by 3 basic lensing parameters of $(t_0, u_0, \te)$, which denote the time of the source 
star's closest approach to the lens, the lens-source separation (impact parameters) at that time, 
and the Einstein time scale, respectively. The impact parameter is scaled to $\thetae$, and the 
Einstein time scale is defined as the time for the source to transit the Einstein radius.  In the 
planetary and binary-lens cases, the lens system is composed of two lens masses and a single source 
(2L1S), and the inclusion of an extra lens component requires including additional parameters to 
depict the lens binarity. These additional parameters are $(s, q, \alpha)$, and the first two 
parameters denote the projected separation (normalized to $\thetae$) and mass ratio between the 
lens components, respectively, and the last one represents the angle between the source trajectory 
and the binary-lens axis (source trajectory angle).  In the binary-source case, the lens system 
comprises a single lens and two source stars (1L2S), and the addition of the extra source requires 
including three additional parameters $(t_{0,2}, u_{0,2}, q_F)$, which denote the approach time and 
impact parameter of the source companion, and the flux ratio between the two source stars, respectively 
\citep{Hwang2013}.  In all tested models, we include an additional parameter $\rho$, which is defined 
as the ratio of the angular source radius to $\thetae$ (normalized source radius), for the consideration 
of finite-source effects, which may affect the lensing light curves of high-magnification events when 
the lens passes over the surface of a source or a central caustic. In the 1L2S model, there are two 
source stars, and we denote the normalized radius of the secondary source as $\rho_2$.  In the following 
subsections, we explain the detailed procedure of modeling conducted for the individual events.

\begin{figure}[t]
\includegraphics[width=\columnwidth]{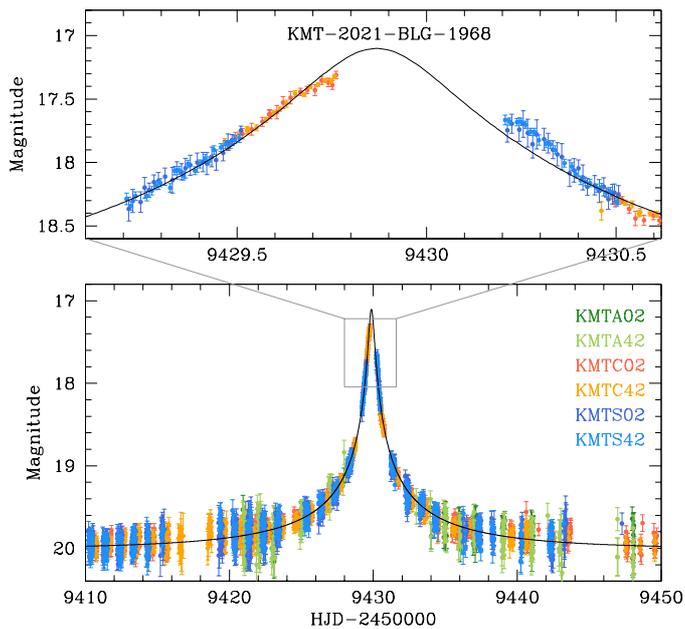}
\caption{
Light curve of KMT-2021-BLG-1968. The lower panel shows the whole view and the upper panel 
shows the enlarged view of the peak region around the anomaly. Drawn over the data is the 
1L1S model curve. 
}
\label{fig:one}
\end{figure}

\begin{figure}[t]
\includegraphics[width=\columnwidth]{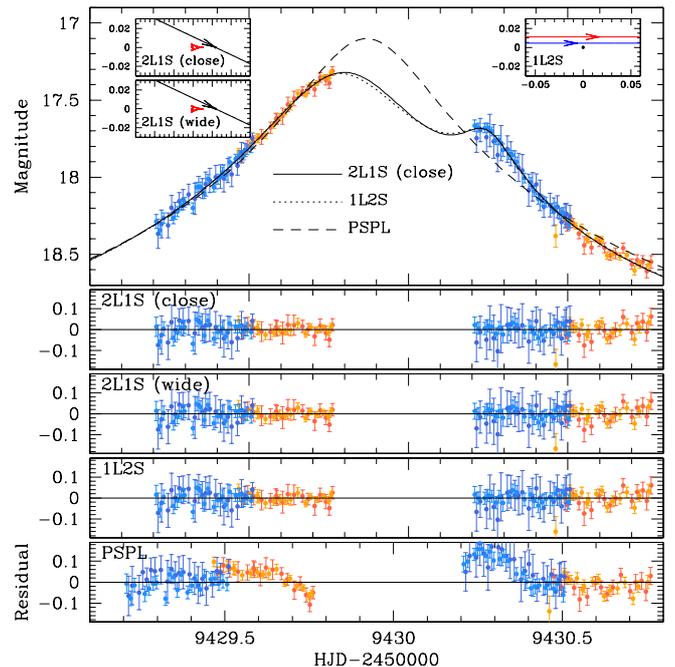}
\caption{
Light curve KMT-2021-BLG-1968 in the region of the anomaly. The curves drawn over the data 
points are 2L1S, 1L2S, and PSPL models and the lower four panels show the residuals from the 
individual models. The three insets in the top panel shows the lens system configurations of 
the close and wide 2L1S solutions and the 1L2S solution. For each 2L1S configuration, the 
cuspy red figure represents the caustic and the arrowed line indicates the source trajectory. 
For the 1L2S configuration, the small filled dot represents the lens position and the arrowed 
lines marked in red and blue represent the trajectories of the primary and secondary source 
stars, respectively.
}
\label{fig:two}
\end{figure}

\begin{figure}[t]
\includegraphics[width=\columnwidth]{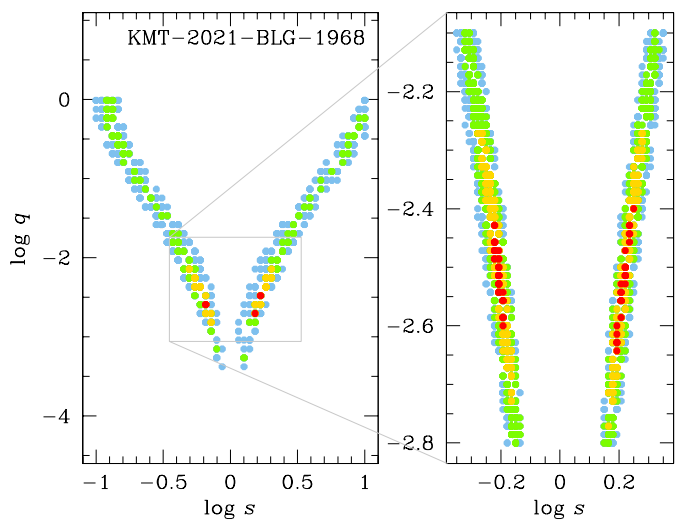}
\caption{
$\Delta\chi^2$ map on the $(\log~s, \log~q)$ plane obtained from the 2L1S modeling of 
KMT-2021-BLG-1968. The right panel shows the enlargement of the map around the local minima.  
Colors are assigned to represent points with $\Delta\chi^2 \leq 1n\sigma$ (red), $\leq 2n\sigma$ 
(yellow), $\leq 3n\sigma$ (green), and $\leq 4n\sigma$ (cyan), where $n=3$.  
}
\label{fig:three}
\end{figure}

\subsection{KMT-2021-BLG-1968 }\label{sec:three-one}

The event KMT-2021-BLG-1968 occurred on a source lying at the Equatorial coordinates 
(RA, DEC)$_{\rm J2000} = (17$:52:25.77, -28:09:05.87), which correspond to the Galactic coordinates 
$(l, b) = (1^\circ\hskip-2pt .443, -0^\circ\hskip-2pt .878)$. The source position corresponds to 
the KMTNet prime fields of BLG02 and BLG42, toward which observations were conducted with a 0.5~hr 
cadence for each field, and a 0.25~hr in combination. The extinction toward the field, $A_I\sim 4.1$, 
was high due to the closeness of the field to the Galactic center. Together with the faintness of 
the source, the measurement of the source baseline was difficult, but the source was registered in 
the Dark Energy Camera (DECam) catalog with an $i$-band baseline magnitude of $i_{\rm base}=21.83$. 
The KMTNet alert of the event was issued on 2021 August 2, which corresponds to the abridged 
heliocentric Julian data of ${\rm HJD}^\prime \equiv {\rm HJD}-2450000 = 9428$. The event reached 
its peak a day after the alert at ${\rm HJD}^\prime =9429.8$ with a peak magnitude of $I_{\rm peak}
\sim 17.4$. The event was observed solely by the KMTNet survey without any followup observations.

The light curve of KMT-2021-BLG-1968 is displayed in Figure~\ref{fig:one}, in which the lower panel 
shows the whole view and the upper panel shows the peak region of the light curve. The curve drawn
over the data points is a point-source point-lens (PSPL) model.  It shows that a short-term
anomaly with about 1~day duration occurred near the peak of the light curve. The residual from
the PSPL model presented in the bottom panel of Figure~\ref{fig:two} shows that negative and positive
deviations occurred in the rising and falling parts, respectively. The time gap between the anomaly
regions covered by the KMTC and KMTS data sets corresponds to the night time of the KMTA site, but
observations at the Australian site could not be done due to bad weather.

For the interpretation of the anomaly, we first conducted a 2L1S modeling of the light curve.  
Figure~\ref{fig:three} shows the $\Delta\chi^2$ map obtained from the grid searches for the 
binary lens parameters $s$ and $q$. The map shows a unique pair of local minima at $(\log s, 
\log q)\sim (\pm 0.2, -2.5)$ resulting from the close-wide degeneracy \citep{Dominik1999, 
An2005}. In Table~\ref{table:one}, we list the full lensing parameters of the close ($s<1$) 
and wide ($s>1$) 2L1S solutions obtained after refining the solutions by allowing all 
parameters to vary. The degeneracy between the solutions is very severe with $\Delta\chi^2  
< 1$. For both solutions, the estimated mass ratios between the lens components, $q \sim 3.1 
\times 10^{-3}$ , are very small, indicating that the companion to the lens is a planet according 
to the 2L1S interpretation.  The model curve (solid curve) of the close planetary solution is 
drawn over the data points in the top panel of Figure~\ref{fig:two}, and the residuals of both 
the close and wide solutions are presented in the lower panels.  The lens system configurations 
of the close and wide 2L1S solutions are presented in the left two insets of the top panel of 
Figure~\ref{fig:two}. It shows that the anomaly was produced by the source passage through the 
anomaly region extending from the protruding cusp of the central caustic induced by a planetary 
companion. The measured event time scale is $\te \sim 20$~days. The normalized source radius 
cannot be accurately measured due to the non-caustic-crossing nature of the anomaly, and only 
the upper limit $\rho_{\rm max}\sim 4\times 10^{-3}$ is constrained.

\begin{figure}[t]
\includegraphics[width=\columnwidth]{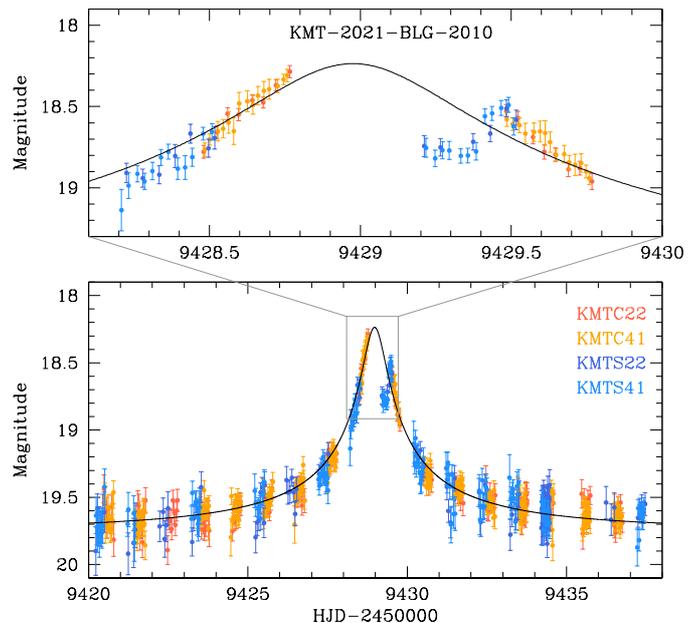}
\caption{
Light curve of KMT-2021-BLG-2010. The layout and scheme of the plots are same as those of
Fig.~\ref{fig:one}.
}
\label{fig:four}
\end{figure}

We check the feasibility of explaining the anomaly with the introduction of a source companion
by additionally conducting a 1L2S modeling. The lensing parameters of the best-fit 1L2S solution
are listed in Table~\ref{table:one}. The fact that $q_F\sim 0.12$ indicates that the secondary 
source is fainter than the primary source, and the fact that $t_0<t_{0,2}$ and $u_0<u_{0,2}$ 
indicates the secondary source trailed the primary source and approached closer to the lens 
than the primary source. The model curve (dotted curve) and the residual of the binary-source 
solution together with the configuration of the 1L2S lens system are presented in Figure~\ref{fig:one}. 
In principle, binary source models sometimes show measurable color-dependent effects, for example, 
MOA-2012-BLG-486 \citep{Hwang2013}.  However, because of the high extinction, the event is too 
faint in the $V$-band to carry out such tests.

\begin{table}[t]
\small
\caption{Model parameters of KMT-2021-BLG-2010 \label{table:two}}
\begin{tabular*}{\columnwidth}{@{\extracolsep{\fill}}llccc}
\hline\hline
\multicolumn{1}{c}{Parameter}    &
\multicolumn{1}{c}{Close}       &
\multicolumn{1}{c}{Wide}      \\
\hline
$\chi^2$/dof            &  $3817.5/3811        $   &  $3817.8/3811        $    \\
$t_{0}$ (HJD$^\prime$)  &  $9428.960 \pm 0.007 $   &  $9428.962 \pm 0.008 $    \\
$u_{0}$                 &  $0.017 \pm 0.003    $   &  $0.017 \pm 0.003    $    \\
$\te$ (days)            &  $15.10 \pm 1.69     $   &  $15.18 \pm 1.67     $    \\
$s$                     &  $0.845 \pm 0.024    $   &  $1.140 \pm 0.036    $    \\
$q$ (10$^{-3}$)         &  $2.77 \pm 0.88      $   &  $2.82 \pm 0.98      $    \\
$\alpha$ (rad)          &  $0.925 \pm 0.084    $   &  $0.918 \pm 0.093    $    \\
$\rho_1$ (10$^{-3}$)    &  $< 5                $   &  $< 5                $    \\ 
\hline
\end{tabular*}
\end{table}

We find that the observed anomaly in the lensing light curve is almost equally well explained 
with the planetary and binary-source interpretations. From the comparison of the fits, it is 
found that the planetary solution is preferred over the binary-source solution by a mere $\Delta
\chi^2=3.0$. We note that resolving the degeneracy between the planetary and binary-source 
solutions would be difficult even if the gap between the KMTC and KMTS data sets had been 
covered by the KMTA data because the two models in the gap region are almost identical.

\subsection{KMT-2021-BLG-2010 }\label{sec:three-two}

The lensing event KMT-2021-BLG-2010 occurred on a source lying at $({\rm RA}, {\rm 
DEC})_{\rm J2000} = ($17:52:02.04, -32:16:11.10), $(l, b) = (-2^\circ\hskip-2pt .148, 
-2^\circ\hskip-2pt .899)$, toward which the $I$-band extinction is $A_I\sim 2.1$. The 
source lies in the KMTNet prime field of BLG41 and the sub-prime field BLG22, toward which 
the combined observational cadence is 0.33 hr.  The event was observed solely by the KMTNet 
group, who issued the alert of the event on 2021 August 03 (${\rm HJD}^\prime=9429$) at 
around the time when the event reached its peak.

\begin{figure}[t]
\includegraphics[width=\columnwidth]{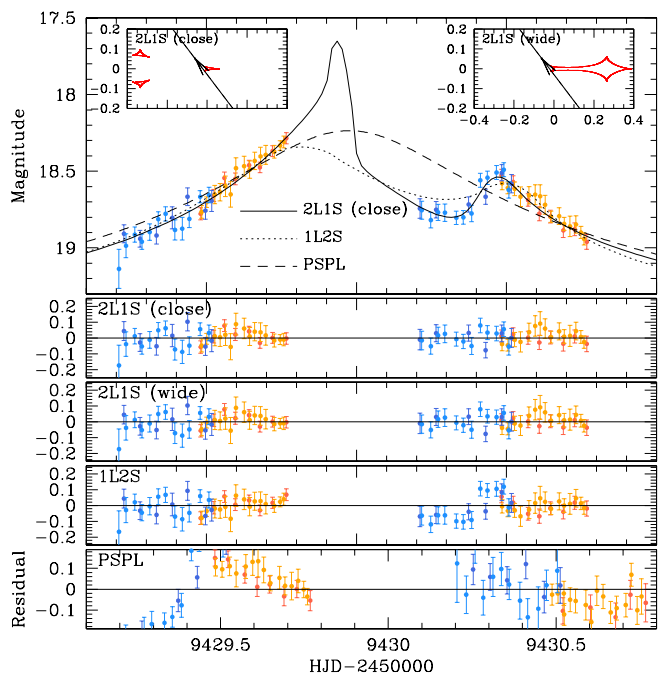}
\caption{
Zoom of the central anomaly region of KMT-2021-BLG-2010. The lower 4 panels show the residuals 
from the close and wide 2L1S, 1L2S, and PSPL models. The two insets in the top panel show the 
lens-system configurations of the close and wide 2L1S solutions. 
}
\label{fig:five}
\end{figure}

In Figure~\ref{fig:four}, we present the light curve of KMT-2021-BLG-2010. The light curve 
is constructed with the use of only the KMTC and KMTS data sets because the photometry quality 
of the KMTA data set is not good. It is found that the light curve exhibits an anomaly appearing 
around the peak with an approximate duration of 1 day, and the deviation from the PSPL model 
appears both in the rising and falling parts as shown in the PSPL residual presented in the 
bottom panel of Figure~\ref{fig:five}. Similar to the case of KMT-2021-BLG-1968, the coverage 
of the anomaly is incomplete due to the absence of the KMTA data set\footnote{Two KMTA data 
points were actually taken during the anomaly, but they proved to be unusable due to bad 
observing conditions.}.

\begin{figure}[t]
\includegraphics[width=\columnwidth]{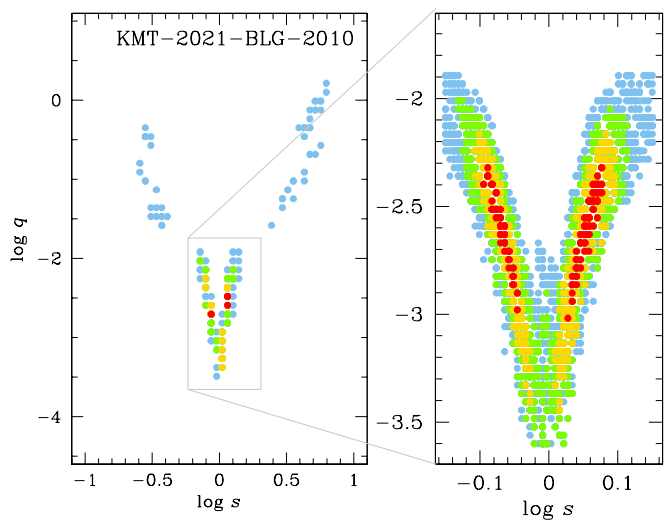}
\caption{
$\Delta\chi^2$ map from the 2L1S modeling of KMT-2021-BLG-2010. Notations and color scheme 
of the plots are same as those of Fig.~\ref{fig:three}, except that $n=2$.
}
\label{fig:six}
\end{figure}

We find that the anomaly of the event is uniquely explained by a pair of planetary models. 
In Figure~\ref{fig:six}, we present the $\Delta\chi^2$ map obtained from the grid searches 
for the binary parameters $s$ and $q$. The estimated binary parameters are $(\log~s, 
\log~q)\sim (\pm 0.06, -2.6)$, indicating that the anomaly was produced by a planetary 
companion and the two solutions result from the close-wide degeneracy. The final lensing 
parameters of the close and wide planetary solutions are listed in Table~\ref{table:two}, 
and the model curve of the close 2L1S solution is drawn over the data points in the top 
panel of Figure~\ref{fig:five}, and the residuals from both the close and wide 2L1S 
solutions are shown in the lower panels. We also present the lens system configurations of 
the close and wide 2L1S solutions in the insets of the top panel. According to the solutions, 
the anomaly was produced by the source passage through the negative deviation region formed 
in the back-end region of the central caustic induced by a planet. The source did not cross 
the caustic, and thus only the upper limit of the normalized source radius, $\rho_{\rm max}
\sim 5\times 10^{-3}$, is constrained.

We find that the planetary and binary-source models can be distinguished with a high confidence
level of $\Delta\chi^2 =62.7$. The model curve and the residual of the best-fit 1L2S solution 
are presented in Figure~\ref{fig:five}, showing that the model results in a poor fit especially 
in the region $9429.2 \lesssim {\rm HJD}^\prime \lesssim  9429.5$, that was covered by the KMTS 
data set.

\begin{table*}[t]
\small
\caption{Model parameters of the solutions "A" and "B" of KMT-2022-BLG-0371 \label{table:three}}
\begin{tabular}{lllll}
\hline\hline
\multicolumn{1}{c}{Parameter}      &
\multicolumn{1}{c}{Sol A, close}   &
\multicolumn{1}{c}{Sol A, wide}    &
\multicolumn{1}{c}{Sol B, close}   &
\multicolumn{1}{c}{Sol B, wide}    \\
\hline
$\chi^2$/dof            &  $1469.2/1487       $  &  $1469.9/1487       $   &  $1469.0/1487       $  &   $1469.4/1487       $    \\
$t_{0}$ (HJD$^\prime$)  &  $9689.419 \pm 0.002$  &  $9689.420 \pm 0.002$   &  $9689.335 \pm 0.003$  &   $9689.335 \pm 0.003$    \\
$u_{0}$ (10$^{-3}$)     &  $2.539 \pm 0.045   $  &  $2.658 \pm 0.111   $   &  $2.089 \pm 0.076   $  &   $2.156 \pm 0.081   $    \\
$\te$ (days)            &  $94.89 \pm1.91     $  &  $89.33 \pm 4.12    $   &  $82.43 \pm 2.56    $  &   $79.93 \pm 3.15    $    \\
$s$                     &  $0.123 \pm 0.018   $  &  $9.689 \pm 1.211   $   &  $0.310 \pm 0.014   $  &   $3.299 \pm 0.063   $    \\
$q$ (10$^{-3}$)         &  $46.09 \pm 37.69   $  &  $70.26 \pm 21.74   $   &  $18.44 \pm 1.90    $  &   $19.82 \pm 1.44    $    \\
$\alpha$ (rad)          &  $2.794 \pm 0.042   $  &  $2.768 \pm 0.049   $   &  $4.918 \pm 0.014   $  &   $4.914 \pm 0.010   $    \\
$\rho$   (10$^{-3}$)    &  $2.53 \pm 0.12     $  &  $2.59 \pm 0.18     $   &  $2.25 \pm 0.08     $  &   $2.33 \pm 0.09     $    \\ 
$\pien$                 &  $-0.74 \pm 0.31    $  &  $0.84 \pm 0.57     $   &  $-0.26 \pm 0.66    $  &   $0.34 \pm 0.60     $    \\ 
$\piee$                 &  $-0.18 \pm 0.08    $  &  $0.04 \pm 0.10     $   &  $-0.04 \pm 0.12    $  &   $0.08 \pm 0.11     $    \\ 
\hline                   
\end{tabular}
\end{table*}

\begin{table*}[t]
\small
\caption{Model parameters of the solutions "C" and "D" of KMT-2022-BLG-0371 \label{table:four}}
\begin{tabular}{lllll}
\hline\hline
\multicolumn{1}{c}{Parameter}      &
\multicolumn{1}{c}{Sol C, close}   &
\multicolumn{1}{c}{Sol C, wide}    &
\multicolumn{1}{c}{Sol D, close}   &
\multicolumn{1}{c}{Sol D, wide}    \\
\hline
$\chi^2$/dof            &  $1470.7/1487       $  &  $1470.7/1487       $   &  $1461.3/1487       $   &   $1462.3/1487       $    \\
$t_{0}$ (HJD$^\prime$)  &  $9689.418 \pm 0.001$  &  $9689.418 \pm 0.001$   &  $9689.363 \pm 0.005$   &   $9689.365 \pm 0.007$    \\
$u_{0}$ (10$^{-3}$)     &  $2.887 \pm 0.124   $  &  $3.024 \pm 0.092   $   &  $1.709 \pm 0.153   $   &   $1.869 \pm 0.164   $    \\
$\te$ (days)            &  $83.328 \pm 3.13   $  &  $80.19 \pm 2.19    $   &  $103.76 \pm 7.56   $   &   $95.08 \pm 7.89    $    \\
$s$                     &  $0.624 \pm 0.077   $  &  $1.714 \pm 0.055   $   &  $0.940 \pm 0.005   $   &   $1.065 \pm 0.006   $    \\
$q$ (10$^{-3}$)         &  $0.72 \pm 0.45     $  &  $0.95 \pm 0.10     $   &  $0.39 \pm 0.05     $   &   $0.42 \pm 0.06     $    \\
$\alpha$ (rad)          &  $4.549 \pm 0.025   $  &  $4.571 \pm 0.026   $   &  $5.311 \pm 0.063   $   &   $5.287 \pm 0.086   $    \\
$\rho$   (10$^{-3}$)    &  $3.00 \pm 0.16     $  &  $3.17 \pm 0.11     $   &  $0.79 \pm 0.10     $   &   $0.89 \pm 0.11     $    \\ 
$\pien$                 &  $0.19 \pm 0.63     $  &  $-0.09 \pm 0.58    $   &  $1.05 \pm 0.68     $   &   $0.86 \pm 0.67     $    \\ 
$\piee$                 &  $0.03 \pm 0.11     $  &  $-0.03 \pm 0.12    $   &  $0.16 \pm 0.127    $   &   $0.10 \pm 0.12     $    \\ 
\hline                   
\end{tabular}
\end{table*}

\subsection{KMT-2022-BLG-0371}\label{sec:three-three}

\begin{figure}[t]
\includegraphics[width=\columnwidth]{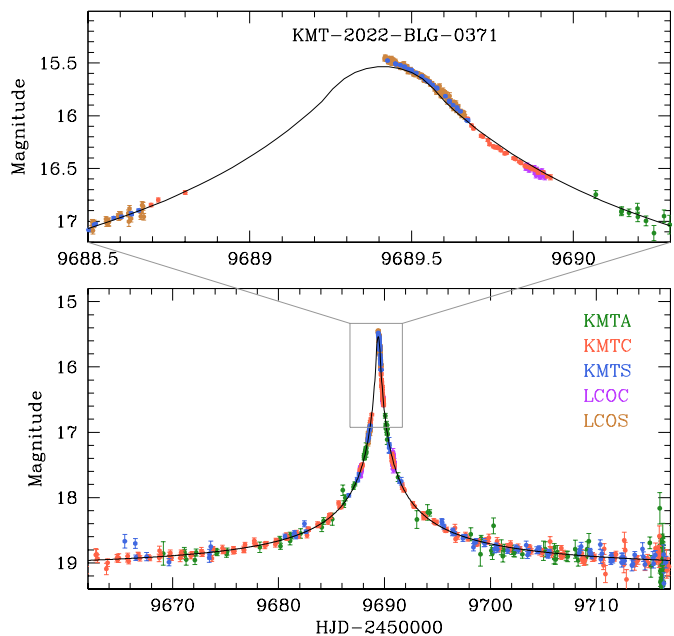}
\caption{
Light curve of KMT-2022-BLG-0371.
}
\label{fig:seven}
\end{figure}

The source of the lensing event KMT-2022-BLG-0371 lies at $({\rm RA}, {\rm DEC})_{\rm J2000} 
=($17:41:26.86, -34:41:55.21), $(l, b)=(-5^\circ\hskip-2pt .372, -2^\circ\hskip-2pt .267)$.  
The $I$-band extinction toward the field is $A_I=2.18$.  The event was found on 2022 April 
11 (${\rm HJD}^\prime=9680$), when the source became brighter than the baseline magnitude of 
$I_{\rm base}=19.72$ by 0.45~mag. The source location corresponds to the KMTNet sub-prime 
field BLG37, toward which observations were carried out with a 2.5~hr cadence. The event 
reached its peak at ${\rm HJD}^\prime =9689.4$ with a very a high magnification of $A_{\rm max} 
\sim 570$.  Four days before the event reached the peak (${\rm HJD}^\prime  = 9685.27$), a 
high-magnification alert was issued by the KMTNet HighMagFinder system \citep{Yang2022}.  In 
response to this alert, two days before the peak, follow-up observations were conducted using 
the LCOC and LCOS telescopes and the cadence of KMTNet observations was increased.  The 
magnification of the source flux lasted more than 100 days, which comprises a significant 
fraction of the Earth's orbital period, that is, 1~yr, and thus we consider microlens-parallax 
effects in modeling the lensing light curve.

\begin{figure}[t]
\includegraphics[width=\columnwidth]{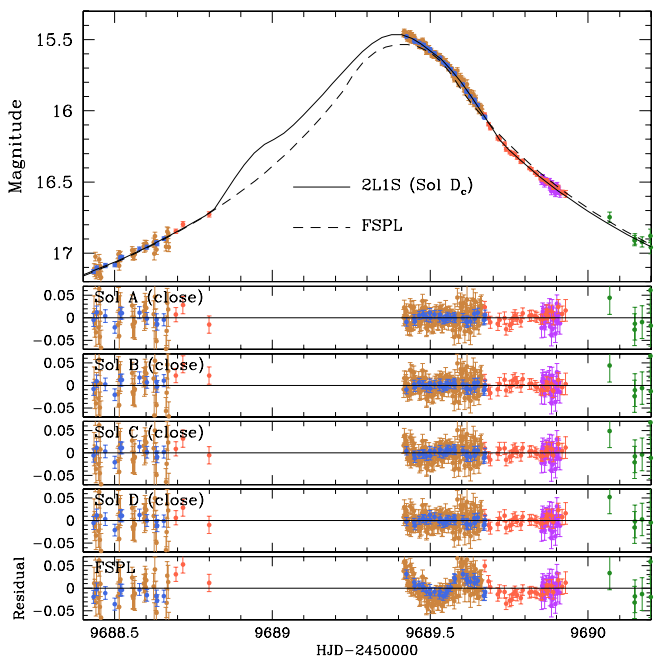}
\caption{
Zoom of the central anomaly region of KMT-2022-BLG-0371. The lower 5 panels show the residuals 
from the 4 degenerate 2L1S (close A, B, C, and D solutions) and FSPL models. 
}
\label{fig:eight}
\end{figure}

\begin{figure}[t]
\includegraphics[width=\columnwidth]{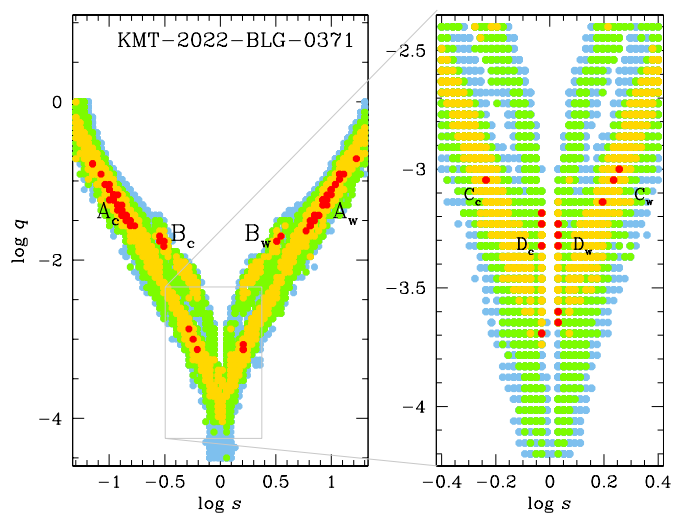}
\caption{
$\Delta\chi^2$ map constructed from the 2L1S modeling of KMT-2022-BLG-0371. Notations and 
color scheme of the plots are same as those of Fig.~\ref{fig:three}, except that $n=2$. 
The locations of the 4 sets of local solutions are marked by "A", "B", "C", and "D", and the 
subscripts "c" and "w" denote the close and wide solutions, respectively.
}
\label{fig:nine}
\end{figure}

\begin{table*}[t]
\small
\caption{Proper motion probability for the 4 local solutions of KMT-2022-BLG-0371 \label{table:five}}
\begin{tabular}{lllllll}
\hline\hline
\multicolumn{1}{c}{Solution}    &
\multicolumn{1}{c}{$\rho$ (10$^{-3}$)}   &
\multicolumn{1}{c}{$\theta_*$ ($\mu$as)}    &
\multicolumn{1}{c}{$\thetae$ (mas)}   &
\multicolumn{1}{c}{$\te$ (day)}   &
\multicolumn{1}{c}{$\mu_{\rm obs}$ (mas/yr)}   &
\multicolumn{1}{c}{$p(\mu<\mu_{\rm obs})$}    \\
\hline
Sol~A  & 2.53  &  0.337  &  0.133   &  95   &  0.51   &  $7  \times 10^{-3}$  \\
Sol~B  & 2.25  &  0.370  &  0.164   &  82   &  0.73   &  $15 \times 10^{-3}$  \\
Sol~C  & 3.00  &  0.367  &  0.122   &  83   &  0.54   &  $8  \times 10^{-3}$  \\
Sol~D  & 0.79  &  0.326  &  0.419   &  104  &  1.45   &  $59 \times 10^{-3}$  \\
\hline                   
\end{tabular}
\end{table*}

The light curve of KMT-2022-BLG-0371 as constructed from the combination of the survey and 
followup data is presented in Figure~\ref{fig:seven}, in which the curve drawn over the data 
is a finite-source point-lens (FSPL) model. It shows that the peak region of the light curve 
exhibits an anomaly that lasted for about 1~day.  While the falling side of the anomaly was 
densely covered by the combination of the KMTS and KMTC survey and the LCOS followup data, 
the rising part of the peak region, which corresponds to the night time in Australia, was not 
covered because site containing KMTA and LCOA was clouded out.  Likewise, it was not possible 
to observe from New Zealand due to clouds.  Despite the partial coverage, the peak region of 
the light curve clearly displays an anomaly feature with a 0.05~mag deviation level as shown 
in the FSPL residual presented in the bottom panel of Figure~\ref{fig:eight}.  Without the 
increased cadence of observations over the peak, the standard KMTNet cadence of 2.5~hr 
($\sim 0.1$~days) for the BLG37 field would not have been sufficient to accurately characterize 
the anomaly.

\begin{figure}[t]
\includegraphics[width=\columnwidth]{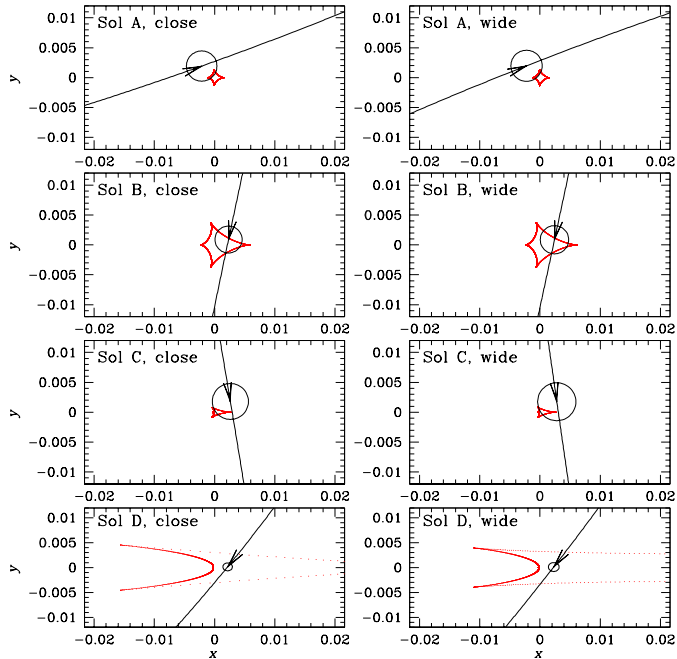}
\caption{
Lens system configurations of the four pairs of the local solutions of KMT-2022-BLG-0371. In 
each panel, the empty circle on the source trajectory represent the source size in units of 
the Einstein radius as given by the tick marks. 
}
\label{fig:ten}
\end{figure}

\begin{figure}[t]
\includegraphics[width=\columnwidth]{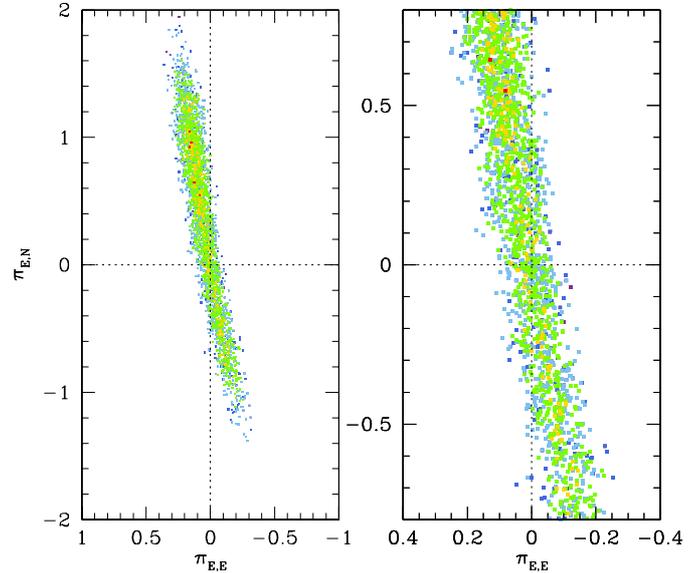}
\caption{
Scatter plot of points in the MCMC chain on the $\pien$--$\piee$ parameter plane of KMT-2022BLG-0371.
The right panel shows the enlargement of the region around the origin. Color scheme of the plots 
are same as those of Fig.~\ref{fig:three}, except that $n=1$. 
}
\label{fig:eleven}
\end{figure}

We find that the anomaly in the lensing light curve is described by multiple sets of local solutions 
with different combinations of $\log~s$ and $\log~q$, as shown in the $\Delta\chi^2$ map presented 
in Figure~\ref{fig:nine}. We identify 4 close-wide pairs of solutions, which we designate as "Sol~A", 
"Sol~B", "Sol~C" and "Sol~D". The refined lensing parameters of the individual solutions 
are presented in Tables~\ref{table:four} (for solutions A and B) and \ref{table:five} (for solutions 
C and D). For the solutions A and B, the mass ratio between the lens components are $q_{\rm A}\sim 
0.05$ and $q_{\rm B}\sim 0.02$, respectively, and thus the companion would likely be a brown dwarf 
according to these solutions. On the other hand, the mass ratios of the solutions C and D are 
$q_{\rm C}\sim 0.8\times 10^{-3}$ and $q_{\rm D}\sim 0.4\times 10^{-3}$, respectively, and thus 
the companion is a planet according to these solutions. In Figure~\ref{fig:eight}, we present the 
model curve of the best-fit 2L1S model (close Sol~D solution) and the close model residuals of all 
local solutions. In Figure~\ref{fig:ten}, we present the lens system configurations of the individual 
local solutions.  We reject the 1L2S interpretation of the anomaly because the 1L2S model yields a 
poorer fit than the 2L1S solution by $\Delta\chi^2=176.6$.

We find that the solution~D is strongly favored over the other 2L1S solutions by a combination of 
two arguments.  First, the solution~D yields a better fit by $\Delta\chi^2 =7.8$, 7.7, and 9.5 than 
the solutions A, B, and C, respectively.  Second, the solution~D results in a higher probability of 
the relative lens-source motion $\mu$ than those of the other solutions. According to Eq.~(22) of 
\citet{Gould2022}, the relative probability for a proper motion of a lensing event to be less than an 
observed value $\mu_{\rm obs}$ is $p(\mu<\mu_{\rm obs}) = (\mu_{\rm obs}/6~{\rm mas}~{\rm yr}^{-1})^2$ 
in the regime of low proper motions.  In Table~\ref{table:five}, we summarize the values of the 
normalized source radius $\rho$, angular source radius $\theta_*$, angular Einstein radius $\thetae=
\theta_*/\rho$, event time scale $\te$, proper motion $\mu = \thetae/\te$, and the resulting probabilities 
$p(\mu<\mu_{\rm obs})$ for the individual local 2L1S solutions.  The detailed procedure of estimating 
$\theta_*$ is described in Sect.~\ref{sec:four}. From the comparison of the probabilities, it is found 
that the solution~D is at least 4 times more probable than any of the other solutions. While each of 
the two arguments based on $\chi^2$ and $\mu$ is not compelling by itself, together they strongly favor 
the solution~D, although the other solutions are not completely ruled out.  We note that direct measurement 
of the proper motion by resolving the lens and source from future high-resolution followup observations 
can ultimately lift the degeneracy among the local solutions.

\begin{table*}[t]
\small
\caption{Model parameters of KMT-2022-BLG-1013 \label{table:six}}
\begin{tabular}{lllll}
\hline\hline
\multicolumn{1}{c}{Parameter}        &
\multicolumn{1}{c}{Sol~A$_{\rm c}$}  &
\multicolumn{1}{c}{Sol~A$_{\rm w}$}  &
\multicolumn{1}{c}{Sol B}            &
\multicolumn{1}{c}{Sol C}            \\
\hline
$\chi^2$/dof            &  $1702.7/2047       $  &  $1704.3/2047       $   &  $1688.9/2047       $   &   $1659.9/2047       $    \\
$t_{0}$ (HJD$^\prime$)  &  $9737.168 \pm 0.002$  &  $9737.164 \pm 0.003$   &  $9737.197 \pm 0.003$   &   $9737.145 \pm 0.002$    \\
$u_{0}$ (10$^{-3}$)     &  $7.60 \pm 0.36     $  &  $6.96 \pm 0.33     $   &  $4.02 \pm 0.23     $   &   $3.58 \pm 0.16     $    \\
$\te$ (days)            &  $22.95 \pm 0.97    $  &  $24.36 \pm 1.02    $   &  $36.69 \pm 1.61    $   &   $40.17 \pm 1.75    $    \\
$s$                     &  $0.518 \pm 0.011   $  &  $1.908 \pm 0.043   $   &  $0.968 \pm 0.002   $   &   $1.072 \pm 0.002   $    \\
$q$ (10$^{-3}$)         &  $14.08 \pm 0.61    $  &  $13.21 \pm 0.59    $   &  $0.23 \pm 0.02     $   &   $1.60 \pm 0.14     $    \\
$\alpha$ (rad)          &  $0.593 \pm 0.016   $  &  $0.555 \pm 0.019   $   &  $3.209 \pm 0.004   $   &   $6.027 \pm 0.007   $    \\
$\rho$   (10$^{-3}$)    &  $7.89 \pm 0.35     $  &  $7.16 \pm 0.32     $   &  $4.52 \pm 0.29     $   &   $4.93 \pm 0.24     $    \\ 
\hline                   
\end{tabular}
\end{table*}

It is found that reliable measurements of the parallax parameters $(\pien, \piee)$ are difficult 
despite the long time scale of the event. This is shown in the scatter plot of $\Delta\chi^2$ on 
the $\pien$--$\piee$ plane presented in Figure~\ref{fig:eleven}. The scatter plot shows that the 
parallax solution is not only consistent with a zero-$\pi_{\rm E}$ model but also the uncertainty 
of $\pivec_{\rm E}$, especially the north component, is very big.

\begin{figure}[t]
\includegraphics[width=\columnwidth]{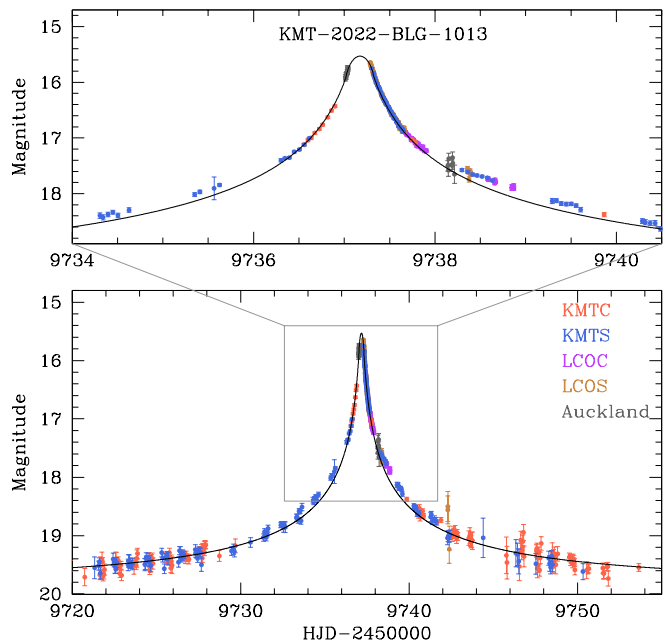}
\caption{
Light curve of KMT-2022-BLG-1013. 
}
\label{fig:twelve}
\end{figure}

\subsection{KMT-2022-BLG-1013}\label{sec:three-four}

The event KMT-2022-BLG-1013, occurred on a bulge star with coordinates $({\rm RA}, 
{\rm DEC})_{\rm J2000} = ($17:38:50.22, -28:21:20.41), $(l, b) = (-0^\circ\hskip-2pt .291, 
1^\circ\hskip-2pt .570)$, was detected in its early stage by the KMTNet survey on 2022 May 
31 (${\rm HJD}^\prime \sim 9729.7$). The baseline magnitude registered in the DECam catalog is 
$i_{\rm base}=21.31$, and the extinction toward the field is $A_I\sim 3.1$. The source was in 
the KMTNet BLG14 field, toward which normal-mode observations were done with a 1.0~hr cadence.  
At ${\rm HJD}^\prime = 9736.89$, the KMTNet HighMagFinder system issued an alert that this 
event was peaking at a high magnification. Then, the peak region of the light curve was 
covered by the KMTS data with an intensified cadence of 0.12~hr during the period of 
$9737.30\leq {\rm HJD}^\prime \leq 9737.65$, and additional data were acquired from followup 
observations conducted using the LCOC, LCOS, and Auckland telescopes.

\begin{figure}[t]
\includegraphics[width=\columnwidth]{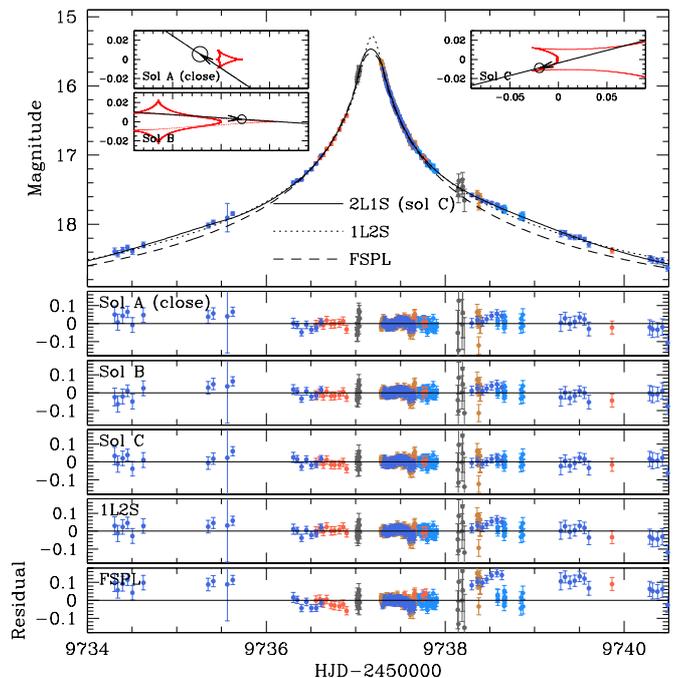}
\caption{
Zoom of the central anomaly region of KMT-2022-BLG-1013. The lower 5 panels show the
residuals from the 3 degenerate 2L1S (solutions A, B, and C), 1L2S, and FSPL models. The 
three insets in the top panel show the 2L1S lens-system configurations of the solutions 
A, B, and C. 
}
\label{fig:thirteen}
\end{figure}

The light curve of KMT-2022-BLG-1013 is presented in Figure~\ref{fig:twelve}. For the 
construction of the light curve, we did not use the KMTA data set due to its poor photometric 
quality.  In any case, there were no KMTA data taken near peak.  The peak region appears to 
exhibit deviations caused by finite-source effects, and thus we first fitted the light curve 
with a FSPL model. It is found that the FSPL model approximately describes the peak, but it 
leaves substantial positive residuals of $\sim 0.1$~mag level in both the rising ($9734
\lesssim {\rm HJD}^\prime \lesssim 9736$) and falling ($9738 \lesssim {\rm HJD}^\prime 
\lesssim 9740$) sides, as shown in the bottom panel of Figure~\ref{fig:thirteen}.

\begin{figure}[t]
\includegraphics[width=\columnwidth]{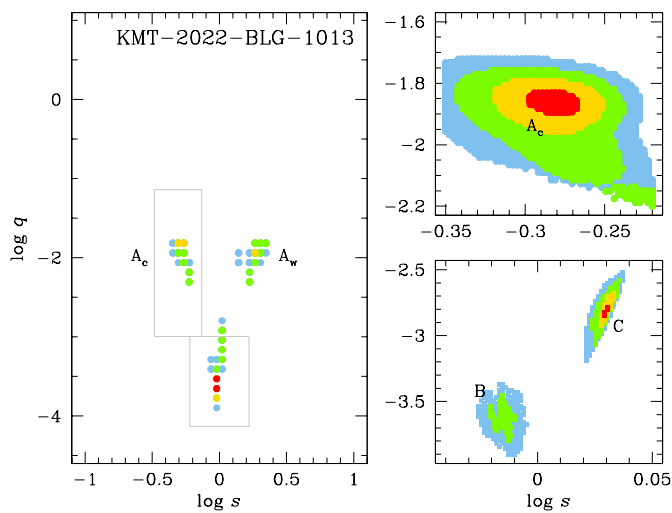}
\caption{
$\Delta\chi^2$ map obtained from the 2L1S modeling
of KMT-2022-BLG-1013. Notations and color scheme are same as those of Fig.~\ref{fig:three}, 
except that $n=2$. 
}
\label{fig:fourteen}
\end{figure}

From the through inspection of the parameter space, it is found that the residuals from the FSPL
model are explained by a unique planetary model. Figure~\ref{fig:fourteen} shows the $\Delta\chi^2$ 
map obtained from the 2L1S grid searches for the binary-lens parameters $s$ and $q$. We identify 
4 local solutions, designated as "Sol~A$_{\rm c}$", "Sol~A$_{\rm w}$", "Sol~B", and "Sol~C", and 
the locations of the individual locals are marked in the $\log~s$--$\log~q$ parameter plane 
presented in Figure~\ref{fig:fourteen}.  We note that the locals Sol~A$_{\rm c}$ and Sol~A$_{\rm w}$ 
are the pair of solutions resulting from the close-wide degeneracy.  In Table~\ref{table:six}, we 
list the refined lensing parameters of the individual local solutions, and the corresponding 
lens-system configurations are presented in the insets of the top panel of Figure~\ref{fig:thirteen}.
We note that the Sol~A$_{\rm c}$ and Sol~A$_{\rm w}$ solutions result in similar configurations, 
and thus we present the configuration of the Sol~A$_{\rm c}$ solution as a representative one.  The 
residuals from the individual models are presented in the lower 4 panels of Figure~\ref{fig:thirteen}. 
From the comparison of the residuals and the $\chi^2$ values of the models, it is found that the 
Sol~C model yields a substantially better fit to the data than the other models, by $\Delta\chi^2 
=42.8$, 44.4, and 29.0 with respect to Sol~A$_{\rm c}$, Sol~A$_{\rm w}$, and Sol~B models, 
respectively. We also find that the Sol~C model is preferred over the 1L2S model by $\Delta\chi^2
=50.7$.

\begin{table*}[t]
\small
\caption{Source stars  \label{table:seven}}
\begin{tabular}{lllll}
\hline\hline
\multicolumn{1}{c}{Quantity}            &
\multicolumn{1}{c}{KMT-2021-BLG-1968 }  &
\multicolumn{1}{c}{KMT-2021-BLG-2010}   &
\multicolumn{1}{c}{KMT-2022-BLG-0371}   &
\multicolumn{1}{c}{KMT-2022-BLG-1013}   \\
\hline
$(V-I)_{\rm S}$            & $2.640 \pm 0.097 $   &  $3.063 \pm 0.234  $  &  $2.375 \pm 0.050 $    &   $4.090 \pm 0.136 $     \\
$I_{\rm S}$                & $22.072 \pm 0.186$   &  $22.830 \pm 0.009)$  &  $22.269 \pm 0.006$    &   $21.652 \pm 0.003$     \\
$(V-I, I)_{\rm RGC}$       & $(2.950, 18.500) $   &  $(2.920, 16.900)  $  &  $(2.697, 16.900) $    &   $(4.260, 17.590) $     \\
$(V-I, I)_{{\rm RGC},0}$   & $(1.060, 14.396) $   &  $(1.060, 14.559)  $  &  $1.060, 14.617   $    &   $(1.060, 14.460) $     \\
$(V-I)_{{\rm S},0}$        & $0.750 \pm 0.097,$   &  $1.203 \pm 0.234  $  &  $0.738 \pm 0.050 $    &   $0.890 \pm 0.136 $     \\
$I_{{\rm S},0}$            & $17.968 \pm 0.186$   &  $20.489 \pm 0.009 $  &  $19.987 \pm 0.006$    &   $18.523 \pm 0.003$     \\
Source type                & G6V                  &  K5V                  &  G4V                   &   K2V                    \\
$\theta_*$ ($\mu$as)       & $0.839 \pm 0.101 $   &  $0.445 \pm 0.109  $  &  $0.326 \pm 0.028 $    &   $0.759 \pm 0.116 $     \\ 
\hline                   
\end{tabular}
\end{table*}

According to the best-fit model, the mass ratio between the lens components is $q\sim 1.6\times 
10^{-3}$, indicating that the companion to the lens is a planet. The projected separation, $s\sim 
1.07$, is very close to unity, and thus the planet induces a resonant caustic. The anomaly was 
produced by the successive caustic crossings of the source, which entered the caustic by crossing 
the upper fold of the caustic at ${\rm HJD}^\prime\sim 9734.6$, passed through the inner caustic 
region, and then exited the caustic by going through the strong lower cusp of the caustic at 
${\rm HJD}^\prime \sim 9738.9$. According to the solution, the positive deviation in the rising 
part of the anomaly is explained by the source crossing over the upper fold caustic, and the 
positive deviation in the falling part is explained by the source passage through the positive 
anomaly region extending from the strong caustic cusp.

\section{Source stars and Einstein radii}\label{sec:four}

In this section, we specify the source stars of the events. The main purpose of the source
specification is estimating the angular Einstein radius of the planetary lens event from the
measured normalized source radius by the relation
\begin{equation}
\thetae = {\theta_*\over \rho},
\label{eq1}
\end{equation}
where the angular source radius $\theta_*$ is deduced from the source type. We are able to 
measure the Einstein radii of the two events KMT-2022-BLG-0371 and KMT-2022-BLG-1013, for which 
the lenses are identified as planetary systems and the $\rho$ values are measured. Although the 
lens of KMT-2021-BLG-2010 is a planetary system, the Einstein radius cannot be measured because 
the $\rho$ value is not measured. In the case of KMT-2021-BLG-1968, neither is the lens uniquely 
identified as a planetary system nor is the $\rho$ value measured.  Nevertheless, we specify the 
source stars of all events for fully characterizing the lensing events.

\begin{figure}[t]
\includegraphics[width=\columnwidth]{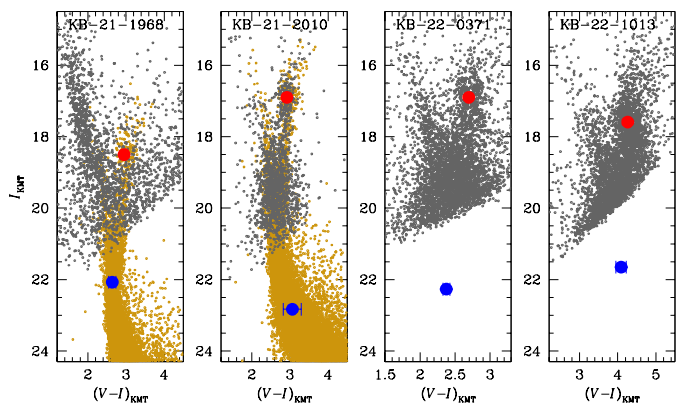}
\caption{
Source locations (filled blue dots) of the events in the instrumental color-magnitude diagrams 
of stars around the source stars. In each panel, the filled red dot denotes the centroid of 
the red giant clump. 
}
\label{fig:fifteen}
\end{figure}

We specified the source star of each event by measuring its de-reddened color $(V-I)_{{\rm S},0}$ 
and magnitude $I_0$ using the \citet{Yoo2004} routine.  Following the routine, we first measured 
the instrumental color and magnitude, $(V-I, I)_{\rm S}$, of the source, placed the source in the 
color-magnitude diagram (CMD) of stars around the source, measured the offset $\Delta(V-I, I)$ of 
the source position from the centroid of the red giant clump (RGC), with $(V-I, I)_{\rm RGC}$, 
in the CMD, and then estimate the de-reddened source color and magnitude as 
\begin{equation}
(V-I, I)_{{\rm S},0} = (V-I, I)_{{\rm RGC},0} + \Delta (V-I, I). 
\label{eq2}
\end{equation}
Here $(V-I, I)_{{\rm RGC},0}$ denote the known values of the de-reddened color and magnitude of 
the RGC centroid from \citet{Bensby2013} and \citet{Nataf2013}, respectively. The instrumental 
source color were measured by regressing the $I$- and $V$-band photometry data with respect to 
the lensing magnification predicted by the model. In the cases of KMT-2021-BLG-1968 and 
KMT-2021-BLG-2010, for which the source stars are very faint, it was difficult to securely 
measure the $V$-band magnitudes. In this case, we utilized the \citet{Holtzman1998} CMD constructed 
from the observations of bulge stars using the Hubble Space Telescope (HST), and interpolate the 
source color from the main-sequence branch on the HST CMD using the measured $I$-band magnitude 
difference between the source and RGC centroid.

\begin{figure}[t]
\includegraphics[width=\columnwidth]{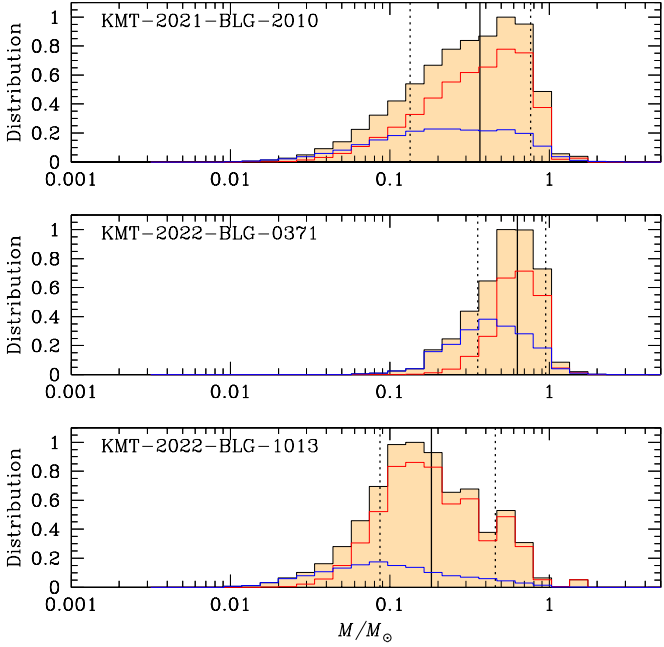}
\caption{
Bayesian posteriors of the mass of the lens system for the planetary events KMT-2021-BLG-2010 
and KMT-2022-BLG-1013. In each panel, the distributions drawn in red and blue represent the 
contributions of the bulge and disk lenses, respectively, and the distribution drawn in black 
is the sum of the contributions by the two lens populations. The solid vertical line indicates 
the median and the dotted vertical lines represent the 16\% and 84\% of the distribution. 
}
\label{fig:sixteen}
\end{figure}

\begin{table}[t]
\small
\caption{Einstein radii and relative lens-source proper motions \label{table:eight}}
\begin{tabular*}{\columnwidth}{@{\extracolsep{\fill}}llccc}
\hline\hline
\multicolumn{1}{c}{Event}           &
\multicolumn{1}{c}{$\thetae$ (mas)} &
\multicolumn{1}{c}{$\mu$ (mas/yr)}  \\
\hline
KMT-2021-BLG-2010  &  $> 0.089$           &  $> 2.15 $          \\
KMT-2022-BLG-0371  &  $0.413 \pm 0.067$   &  $1.45 \pm 0.24$    \\
KMT-2022-BLG-1013  &  $0.154 \pm 0.024$   &  $1.40 \pm 0.22$    \\
\hline
\end{tabular*}
\end{table}

Figure~\ref{fig:fifteen} shows the positions of the source stars and RGC centroids in the CMDs 
of the individual events. In Table~\ref{table:seven}, we list the values of $(V-I, I)_{\rm S}$, 
$(V-I, I)_{{\rm S},0}$, $(V-I, I)_{\rm RGC}$, and $(V-I, I)_{{\rm RGC},0}$ together with the 
stellar types determined from the de-reddened color and magnitude.  For the estimation of the 
angular source radius from $(V-I, I)_{{\rm S},0}$, we first converted the $V-I$ color into $V-K$ 
color, and then deduce $\theta_*$ from the \citet{Kervella2004} relation between $(V-K, V)$ and 
$\theta_*$. The estimated source radii of the individual events are listed in Table~\ref{table:seven}.

With the measured source radius, we estimated the angular Einstein radii using the relation 
in Equation~(\ref{eq1}). We also estimated the relative lens-source proper motion from the 
combination of the Einstein radius and event time scale by
\begin{equation}
\mu = {\thetae\over \te}.
\label{eq3}
\end{equation}
In Table~\ref{table:eight}, we list the $\thetae$ and $\mu$ values of the events for which the 
planetary nature of the lenses are identified.

\begin{figure}[t]
\includegraphics[width=\columnwidth]{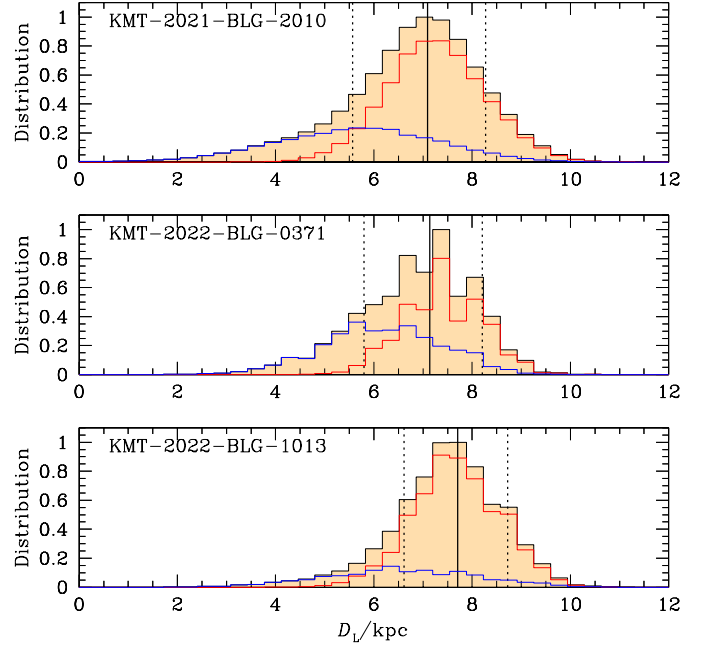}
\caption{
Bayesian posteriors of the distance to the lens system for the events KMT-2021-BLG-2010 and 
KMT-2022-BLG-1013. Notations are same as in Fig.~\ref{fig:sixteen}. 
}
\label{fig:seventeen}
\end{figure}

\section{Physical parameters}\label{sec:five}

In this section, we estimate the physical parameters of the three identified planetary lens 
systems of KMT-2021-BLG-2010L, KMT-2022-BLG-0371L, and KMT-2022-BLG-1013L.  For none of these 
events, the lensing observables $(\te, \thetae, \pie)$ are not fully measured, making it 
difficult to uniquely determine the mass $M$ and distance $\dl$ to the planetary systems 
via the analytic relations 
\begin{equation}
M={\thetae\over \kappa \pie};\qquad
\dl = {{\rm AU} \over \pie\thetae + \pi_{\rm S} }.
\label{eq4}
\end{equation}
Here 
$\kappa=4G/(c^2{\rm AU})$, $\pi_{\rm S}={\rm AU}/\ds$ denotes the parallax of the source, 
and $\ds$ represents the distance to the source.  We, therefore, estimate the physical lens 
parameters by conducting Bayesian analyses based on the measured observables of the individual 
events.

\begin{table*}[t]
\small
\caption{Physical lens parameters  \label{table:nine}}
\begin{tabular}{lrrr}
\hline\hline
\multicolumn{1}{c}{Parameter}          &
\multicolumn{1}{c}{KMT-2021-BLG-2010}  &
\multicolumn{1}{c}{KMT-2022-BLG-0371}  &
\multicolumn{1}{c}{KMT-2022-BLG-1013}  \\
\hline
$M_{\rm p}$ ($M_{\rm J}$)  &  $1.07^{+1.15}_{-0.68}$ & $0.26^{+0.13}_{-0.11}$   &  $0.31^{+0.46}_{-0.16}$  \\  [0.8ex]
$M_{\rm h}$ ($M_\odot$)    &  $0.37^{+0.40}_{0.23} $ & $0.63^{+0.32}_{-0.28}$   &  $0.18^{+0.28}_{-0.10}$  \\  [0.8ex]
$\dl$ (kpc)                &  $7.09^{1.18}_{-1.53} $ & $7.14^{+1.07}_{-1.34}$   &  $7.71^{+1.01}_{-1.09}$  \\  [0.8ex]
$a_\perp$ (AU) (close)     &  $1.79^{+0.30}_{-0.38}$ & $3.02^{+0.45}_{-0.56}$   &  $1.38^{+0.18}_{-0.20}$  \\  [0.8ex]
\hskip32pt     (wide)      &  $2.42^{+0.40}_{-0.52}$ & $3.42^{+0.51}_{-0.64}$   &  --                      \\  [0.8ex]
\hline                   
\end{tabular}
\end{table*}

The Bayesian analysis of each lensing event was carried by conducting a Monte Carlo simulation 
of Galactic lensing events. From the simulation, we produced a large number ($10^7$) of lensing 
events, for which the locations of the lens and source and the relative proper motions were 
assigned based on a prior Galactic model, and the lens masses were allocated based on a prior 
mass function of Galactic objects. In the simulation, we adopted the Galactic model of 
\citet{Jung2021} and the mass function model of \citet{Jung2018}. For each simulated event, we 
computed lensing observables by
\begin{equation}
\te={\thetae\over \mu};\qquad
\thetae = (\kappa M \pi_{\rm rel})^{1/2};\qquad
\pie = {\pi_{\rm rel} \over \thetae},
\label{eq5}
\end{equation}
where the relative parallax is defined as $\pi_{\rm rel} = \pi_{\rm L} - \pi_{\rm S} = {\rm AU}
(1/\dl - 1/\ds)$. We then construct posterior distributions of the physical parameters by imposing 
a weight $w_i=\exp(-\chi^2/2)$ to each simulated event. Here the $\chi^2$ value is computed as 
\begin{equation}
\chi_i^2 = 
\left( {t_{{\rm E},i} -\te \over \sigma_{\te}} \right)^2 + 
\left( {\theta_{{\rm E},i} -\thetae \over \sigma_{\thetae}} \right)^2, 
\label{eq6}
\end{equation}
where $(t_{{\rm E},i}, \theta_{{\rm E},i})$ represent the time scale and Einstein radius of 
each simulated event, and $(\te, \thetae)$ denote the measured values. In the case of 
KMT-2021-BLG-2010, for which only the lower limit of $\thetae$ is constrained, we set $w_i=0$ 
for simulated events with $\theta_{{\rm E},i} < \theta_{\rm E,min}$. Although the constraint 
given by the measured microlens parallax parameters is weak in the case of KMT-2022-BLG-0371, 
we imposed the $\pie$ constraint by adding an additional term 
\begin{equation}
\sum_{j=1}^2
\sum_{k=1}^2
b_{j,k}
(\pi_{{\rm E},j,i}-\pie)
(\pi_{{\rm E},k,i}-\pie)
\label{eq7}
\end{equation}
to the right side of Equation~(\ref{eq6}). Here $b_{jk}$ represents the inverse matrix of 
$\pivec_{\rm E}$, $(\pi_{{\rm E},1}, \pi_{{\rm E},2})_i = (\pi_{{\rm E},N}, \pi_{{\rm E},E})_i$ 
denote the parallax parameters of each simulated event, and $(\pien, \piee)$ are the measured 
parallax parameters.

We present the posteriors of the mass and distance to the three planetary systems in 
Figures~\ref{fig:sixteen} and ~\ref{fig:seventeen}, respectively. In Table~\ref{table:nine}, 
we list the estimated masses of the host, $M_{\rm h}$, and planet, $M_{\rm p}$, distance, $\dl$, 
and projected separation of the planet from the host, $a_\perp =s \dl \thetae$. For each parameter, 
we present the median as a representative value and the uncertainties are estimated as the 16\% 
and 84\% of the posterior distribution. For the events with degenerate close and wide solutions, 
we present a pair of the projected separations resulting from the close and wide solutions. 
According to the estimated planet and host masses, the planetary system KMT-2021-BLG-2010L is 
composed of a Jovian-mass planet and an M-dwarf host, the system KMT-2022-BLG-0371L consists of 
a sub-Jovian-mass planet and a K-dwarf host, and the system KMT-2022-BLG-1013L is composed of a 
sub-Jovian-mass planet and an M-dwarf host.

\section{Summary and conclusion}\label{sec:six}

We conducted analyses of 4 anomalous microlensing events detected in the 2021 and 2022 seasons, 
including KMT-2021-BLG-1968, KMT-2021-BLG-2010, KMT-2022-BLG-0371, and KMT-2022-BLG-1013.  The 
light curves of the events commonly exhibit partially covered short-term central anomalies 
appearing near the highly magnified peaks.  In order to reveal the nature of the anomalies, we 
tested various models that could give rise to the anomalies of the individual events including 
the binary-lens and binary-source interpretations.  Under the 2L1S interpretation, we thoroughly 
inspected the parameter space to check the existence of degenerate solutions, and if they exist, 
we tested the feasibility of resolving the degeneracy.

We found that the anomalies of the two events KMT-2021-BLG-2010 and KMT-2022-BLG-1013 were 
uniquely explained by planetary-lens interpretations, for which the planet-to-host mass ratios 
of $q\sim 2.8\times 10^{-3}$ and $\sim 1.6\times 10^{-3}$, respectively.  For KMT-2022-BLG-0371, 
we found multiple sets of local 2L1S solutions with different combinations of binary parameters, 
but a planetary solution with a mass ratio $q\sim 4\times 10^{-4}$ was strongly preferred over 
the other degenerate solutions based on the $\chi^2$ and relative proper motion arguments, and 
a 1L2S solution was clearly ruled out.  For KMT-2021-BLG-1968, on the other hand, it was found 
that the anomaly could be explained either by a planetary or a binary-source interpretation, 
making it difficult to firmly identify the nature of the anomaly.

From the Bayesian analyses of the identified planetary events, we estimated that the masses of 
the planet and host were
$(M_{\rm p}/M_{\rm J}, M_{\rm h}/M_\odot) = 
 (1.07^{+1.15}_{-0.68}, 0.37^{+0.40}_{-0.23})$, 
$(0.26^{+0.13}_{-0.11}, 0.63^{+0.32}_{-0.28})$, and
$(0.31^{+0.46}_{-0.16}, 0.18^{+0.28}_{-0.10})$ 
for KMT-2021-BLG-2010L, KMT-2022-BLG-0371L, and KMT-2022-BLG-1013L, respectively.

\begin{acknowledgements}
Work by C.H. was supported by the grants of National Research Foundation of Korea 
(2020R1A4A2002885 and 2019R1A2C2085965).
J.C.Y., I.G.S, and S.J.C. acknowledge support from NSF Grant No. AST-2108414.
W.Z. and H.Y. acknowledge the support by the National Science Foundation of
China (Grant No. 12133005). 
Y.S. acknowledges support from BSF Grant No. 2020740.
This research has made use of the KMTNet system operated by the Korea Astronomy and Space 
Science Institute (KASI) at three host sites of CTIO in Chile, SAAO in South Africa, and 
SSO in Australia. Data transfer from the host site to KASI was supported by the Korea 
Research Environment Open NETwork (KREONET). 
W.Zang, J.Z., H.Y., S.M., and W.Zhu acknowledge support by the National Natural Science
Foundation of China (Grant No. 12133005). W.Zang acknowledges the support from the
Harvard-Smithsonian Center for Astrophysics through the CfA Fellowship. This research uses data
obtained through the Telescope Access Program (TAP), which has been funded by the TAP
member institutes. W.Zhu acknowledges the science research grants from the China Manned Space
Project with No.\ CMS-CSST-2021-A11. The authors acknowledge the Tsinghua Astrophysics
High-Performance Computing platform at Tsinghua University for providing computational and data
storage resources that have contributed to the research results reported within this paper.
\end{acknowledgements}

\end{document}